\documentclass[journal]{IEEEtran}

\usepackage{cite}
\usepackage{amsmath,amssymb,amsfonts}
\usepackage{algorithmic}
\usepackage{graphicx}
\usepackage{textcomp}
\usepackage{xcolor}
\usepackage{booktabs}
\usepackage{multirow}
\usepackage{url}
\usepackage{algorithm}
\usepackage{flushend}
\usepackage{graphicx}
\usepackage{hyperref}
\usepackage{subcaption} 

\ifCLASSINFOpdf
\else
\fi

\usepackage{amsmath}
\usepackage{algorithmic}

\begin{document}

\title{E-THER: A Multimodal Dataset for Empathic AI - Towards Emotional Mismatch Awareness}

\author{Sharjeel~Tahir*,
        Judith~Johnson,
        Jumana~Abu-Khalaf,
        and~Syed~Afaq~Ali~Shah
\thanks{S. Tahir*, J. Abu-Khalaf and A. Shah are with the Centre for AI and ML, Edith Cowan University, Joondalup, Australia - 
e-mail*: s.tahir@ecu.edu.au}
\thanks{J. Johnson is with University of Manchester.}
\thanks{Manuscript received [Date]; revised [Date].}}

\markboth{IEEE Transactions on [Journal Name],~Vol.~X, No.~Y, [Month]~[Year]}%
{Tahir \MakeLowercase{\textit{et al.}}: PCT-Grounded Multimodal Dataset for Therapeutically Congruent VLMs}

\maketitle

\begin{abstract}
A prevalent shortfall among current empathic AI systems is their inability to recognize when verbal expressions may not fully reflect underlying emotional states. This is because the existing datasets, used for the training of these systems, focus on surface-level emotion recognition without addressing the complex verbal-visual incongruence (mismatch) patterns useful for empathic understanding. In this paper, we present E-THER, the first Person-Centered Therapy (PCT)-grounded multimodal dataset with multidimensional annotations for verbal-visual incongruence detection, enabling training of AI systems to advance towards realistic rather than performative empathic capabilities. The annotations included in the dataset are drawn from humanistic approach, i.e., identifying verbal-visual emotional misalignment in client-counsellor interactions - forming a framework for training and evaluating AI on empathy tasks. Additional engagement scores provide behavioral annotations for research applications. Notable gains in empathic and therapeutic conversational qualities are observed in state-of-the-art vision-language models, such as IDEFICS and VideoLLAVA, using evaluation metrics following empathic and therapeutic principles. Empirical findings indicate that our incongruence-trained models outperform general state-of-the-art models in critical traits, such as sustaining therapeutic engagement, minimizing artificial or exaggerated linguistic patterns, and maintaining fidelity to PCT theoretical framework.
\end{abstract}


\section{Introduction}
Empathic dialogue generation is a central challenge for human–AI interaction. While large language models (LLMs) and vision–language models (VLMs) have advanced open-domain conversation, systems still tend to rely on surface regularities in text, producing responses that appear empathic without deeper situational grounding \cite{rashkin2018towards, welivita2024large}. In applied communication settings, this gap is often described as \emph{performative empathy} - language that signals care but does not reflect nuanced understanding \cite{montemayor2022principle}.

A key source of nuance arises from multimodal inconsistency: what people say can diverge from how they present nonverbally. Counseling and communication theories describe such verbal–visual incongruence as diagnostically meaningful \cite{rogers1957necessary, mehrabian1967decoding}. Existing empathy and emotion resources, spanning text-only datasets \cite{buechel2018modeling, lin2019moel} and multimodal corpora \cite{busso2008iemocap, poria2019meld, zhu2023medic, chen2024towards}, provide valuable foundations for response generation and emotion recognition. However, systematic annotation and modeling of verbal–visual incongruence for \emph{empathic response generation} remain comparatively underexplored. In parallel, engagement level (e.g., low vs. high client engagement) is known to shape effective conversational strategies \cite{miller2012motivational}, yet many pipelines treat empathic behaviors as context-invariant despite evidence that engagement awareness can enhance interaction quality \cite{kolomaznik2024role}.

This paper addresses these gaps by introducing the \textbf{Empathic THERapy Conversations (E-THER)} dataset and a modeling–evaluation toolkit centered on incongruence-aware empathic communication. E-THER provides multimodal therapeutic dialogues with systematic annotations of verbal–visual mismatch, guided by Person-Centered Therapy (PCT) constructs. We use PCT as a \emph{theoretical lens} because its core emphasis on empathy, unconditional positive regard, and congruence aligns closely with empathic communication, rather than as a therapeutic protocol to be delivered by AI \cite{rogers1957necessary, watson2014role}. Figure~\ref{fig:verbal_visual} illustrates an example of verbal–visual mismatch and its annotation.

Our main contributions are as follows: (1) \textit{E-THER dataset:} a multimodal dialogue benchmark with PCT-guided annotations (focusing on mismatch between speaker's expressions and words) of verbal–visual incongruence to support modeling beyond surface cues.
(2) \textit{Incongruence-aware training:} methods that encourage models to attend to potential mismatch between what is said and shown, fostering more contextually grounded empathic responses \cite{rogers1961becoming}.
(3) \textit{Aligned evaluation:} an automatic evaluation framework tailored to our annotations, i.e., conversational authenticity, responsive engagement, and alignment with Rogers’ core conditions \cite{rogers1959theory}, designed to better reflect empathic communication quality.

Across experiments with multiple VLMs, these components improve incongruence detection and yield higher ratings on our empathy-aligned metrics compared to strong baselines (Sections \ref{sec:eval}–\ref{section:results}), with ablations isolating the contribution of each component (Section \ref{sec:ablation}).

Our focus is \emph{empathic communication in AI}. We do not claim clinical efficacy or propose AI-delivered therapy. PCT is used to define, annotate, and evaluate empathic behaviors and to highlight safety considerations (e.g., non-directiveness, avoidance of unsolicited advice) relevant to supportive, non-clinical interactions \cite{hardman2019friendly, bozarth1998person, stephenson2024empathicAI}.

The remainder of the paper is organized as follows. Section \ref{sec:related} situates our work within artificial empathy research. Section \ref{sec: dataset} details E-THER dataset - construction, annotation methodology, and validation. Section \ref{sec:train} presents our training procedures for incongruence-aware response generation. Section \ref{sec:eval} reports the evaluation setup, and Section \ref{section:results} provides results and analyses, followed by ablations in Section \ref{sec:ablation}. Section \ref{sec:limit} discusses limitations and future directions, and Section \ref{sec:conc} concludes.

\section{Related Work} \label{sec:related}

\subsection{Empathic AI and Dialogue Systems}

Existing artificial empathy research has focused primarily on generating emotionally appropriate responses in conversational settings. The EmpatheticDialogues dataset \cite{rashkin2018towards} is a leading and comprehensive dataset in the domain that is also publicly available, providing 25,000 conversations grounded in emotional situations. Building upon this foundation, ESConv \cite{liu2021towards} introduced emotional support conversation as a structured task with 1,053 conversation exchanges incorporating eight support strategies grounded in Helping Skills Theory, which draws heavily from PCT. More recently, STICKERCONV \cite{zhang2024stickerconv} presented the first comprehensive multimodal empathetic dialogue (conversations) dataset with 12.9K sessions and visual sticker responses, while EDOS \cite{welivita2021large} contributed a large-scale dataset focused specifically on empathetic response generation. However, these approaches primarily emphasize response generation with limited focus on underlying cognitive processes that characterize empathic understanding.

Incorporating emotional reasoning into empathic response generation through techniques such as emotion-cause recognition \cite{wang2023emotionx} and multi-level empathy modeling \cite{sabour2022computational} has been seen in recent works. Advanced frameworks have emerged including LLM-based empathetic generation \cite{qian2023harnessing} and multi-dimensional evaluation approaches \cite{xu2024multi}. Computational empathy has also been explored in mental health support contexts \cite{sharma2020computational}, demonstrating the potential for AI systems to understand and respond to emotional distress. While these approaches represent important advances, they primarily emphasize response generation over empathic reasoning processes or the ability to detect emotional incongruence.

\subsection{Therapeutic and Empathy Datasets}

The landscape of therapeutic dialogue datasets has expanded significantly, yet opportunities exist to enhance clinical grounding and theoretical grounding. ESConv \cite{liu2021towards} introduced emotional support conversation as a structured task, incorporating eight support strategies. However, these conversations rely on crowdsourced interactions that differ from clinical therapeutic settings. Recent multimodal datasets have begun addressing this limitation: MODMA \cite{cai2022multimodal} provides the first multi-modal open dataset for mental-disorder analysis with 53 participants including both clinically depressed patients and healthy controls, combining EEG and spoken language data. 

Recent multimodal approaches have attempted to address empathy detection in therapeutic contexts. MEDIC \cite{zhu2023medic} provides 771 video clips from counseling sessions with empathy mechanism annotations, while MESC \cite{chen2024towards} extends this to comprehensive multimodal emotional support conversations. Clinical dialogue datasets have also emerged, including MTS-Dialog for doctor-patient encounters \cite{abacha2023mts}.
General emotion recognition datasets, including IEMOCAP \cite{busso2008iemocap} and MELD \cite{poria2019meld}, provide multimodal emotion annotations but focus on classification rather than empathic understanding. These datasets utilize acted scenarios or entertainment content that may not generalize to therapeutic interactions, suggesting value in clinically informed datasets.

\subsection{Multimodal Emotion Recognition and Incongruence Detection}
The integration of visual and textual information for emotion recognition has shown significant promise \cite{zhang2023multimodal, dai2023instructblip}, but existing multimodal approaches  primarily focuses on emotion classification tasks rather than the nuanced detection of emotional misalignment patterns. Alexithymia research demonstrates that individuals can exhibit systematic cross-modal emotional inconsistencies \cite{bagby2004multimodal}. Machine learning approaches have been developed to identify complex emotions in alexithymia-affected individuals \cite{gannouni2022identifying}, highlighting the clinical relevance of emotion discrepancy detection.

Large vision-language models have shown promise for contextual emotion recognition \cite{etesam2024contextual}, while specialized approaches for micro-expression analysis using vision transformers demonstrate effectiveness in detecting subtle emotional cues \cite{wang2025leveraging}. Comprehensive surveys indicate that multimodal emotion recognition with deep learning continues to face challenges in handling cross-modal inconsistencies \cite{ramaswamy2024multimodal}.

Advances in vision-language models demonstrate capacity for understanding complex visual-textual relationships \cite{li2023blip2}, yet these capabilities have not been systematically applied to therapeutic or empathic communication analysis. This represents a research opportunity, given the successful integration of VLMs in various relevant tasks including emotion recognition from multimodal content \cite{gandhi2023multimodal}, mental health assessment through visual and textual cues \cite{yoon2023vlm}, and healthcare communication evaluation \cite{chen2023medical}.

\begin{figure*}[htbp]
    \centering
    \includegraphics[width=\textwidth]{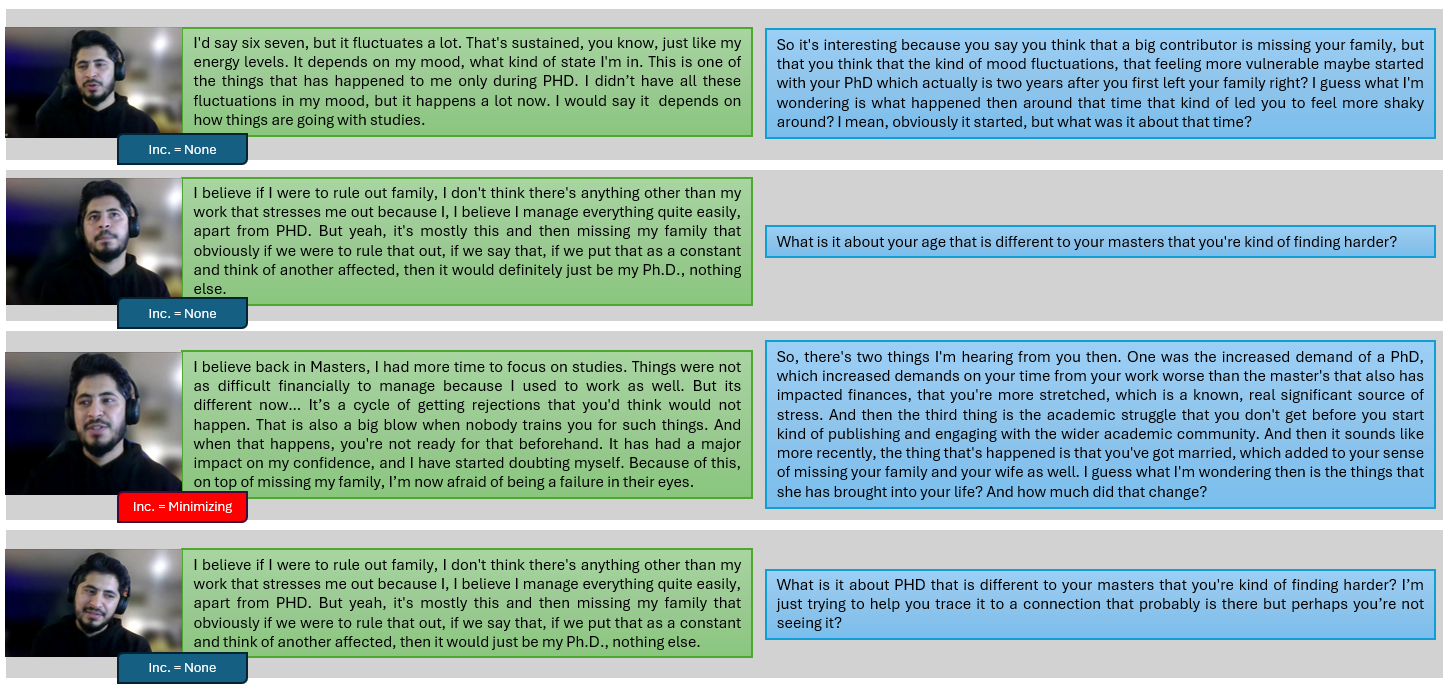}
    \caption{A snippet from one of the recorded conversations that depicts the verbal and visual content of the conversations in the E-THER dataset. It also contains an incongruent example of (minimizing) type where the client verbally expresses emotional distress, while their facial expression remains predominantly neutral - illustrating a cross-modal affective mismatch. Such incongruences are manually annotated (I = 1) against each dialogue pair as such in our dataset to support emotion modeling beyond surface cues. The counsellor's response reflects PCT principles, using non-directive reflection and open inquiry to support client-led emotion discovery. }
    \label{fig:verbal_visual}
\end{figure*}

\subsection{Evaluation Approaches for Empathic Systems}

One of the limitations in current artificial empathy research is the reliance on evaluation metrics with limited prowess in capturing empathic communication nuances~\cite{sharma2023computational}. Traditional approaches use lexical matching (BLEU, ROUGE) or semantic matching with gold-standard over empathic understanding~\cite{xu-jiang-2024-multi, liu2016not}. These word-overlap metrics demonstrate weak or no correlation with human judgments in dialogue evaluation, showing limited effectiveness for the nuanced requirements of empathic communication~\cite{giorgi2023psychological}. Recent empirical studies reveal that BLEU and ROUGE scores show minimal correlation with human assessments of dialogue quality, with correlations often approaching zero in conversational contexts~\cite{liu2016not}. 

More sophisticated evaluation frameworks for empathy in artificial agents have been introduced lately. The Perceived Empathy of Technology Scale (PETS) provides a validated instrument for measuring users' perceptions of AI system empathy \cite{PETS2024}, while research on third-party evaluation demonstrates that AI can be perceived as more compassionate than expert humans in certain contexts \cite{ovsyannikova2025third}. Multi-dimensional evaluation approaches have emerged that assess empathy across cognitive, affective, and behavioral dimensions \cite{xu2024multi}, moving beyond surface-level linguistic analysis. Additionally, specialized algorithms for empathy categorization across conversations have been developed \cite{provence2024algorithms}, offering more nuanced evaluation.

\subsection{Person-Centered Therapy and Computational Applications}

The PCT framework provides established principles for empathic communication through Rogers' core therapeutic conditions: empathy, unconditional positive regard (the therapist’s acceptance and non-judgmental valuing of the client regardless of what is disclosed), and emotional congruence (the consistency between verbal expressions and nonverbal cues, where mismatches may reveal underlying emotions) \cite{rogers1957necessary}. These conditions have been empirically validated across multiple therapeutic modalities and represent fundamental requirements for effective therapeutic relationships \cite{watson2014role}. However, none of the existing artificial empathy approaches have operationalized these principles for empathic AI development.

Recent computational approaches to therapy have focused majorly on Cognitive Behavioral Therapy techniques \cite{fitzpatrick2017delivering} or general mental health support \cite{chancellor2016mental}, with limited attention to PCT-specific empathic communication patterns. The systematic annotation of Rogers' conditions in therapeutic interactions represents a novel contribution to computational psychology research.

While artificial empathy systems show promise in supportive roles, research indicates fundamental distinctions between simulated and therapeutic relationships \cite{shen2024empathy}. Our work positions itself within the domain of supportive AI tools and research applications rather than clinical therapeutic contexts, focusing on advancing computational understanding of empathic communication patterns that characterize effective therapeutic interactions.

\section{Development of E-THER} \label{sec: dataset}

\subsection{Theoretical Foundation and Data Collection}

Our dataset comprises multimodal conversational data from empathic and therapeutic conversations, providing the first source specifically designed for training verbal-visual mismatch-aware empathic AI grounded in established therapeutic theory \cite{hardman2019friendly, kaluzeviciute2020role, mahon2023evidence}. The dataset includes synchronized video, audio, and transcript data from 18 therapeutic sessions conducted by a registered clinical psychologist with 18 participants, totaling approximately 5 hours of interaction data with 1578 dialogue turns after data cleaning.

The sessions were conducted following PCT principles, emphasizing non-directive, empathic communication that supports client self-discovery rather than prescriptive guidance \cite{rogers1957necessary}. This approach provides a theoretically grounded and ethically appropriate framework for AI empathic training, minimizing risks of inappropriate therapeutic boundary crossing while developing empathic understanding capabilities.

\begin{table*}[htbp]
\centering
\caption{Comprehensive Comparison of E-THER with Existing Datasets}
\label{tab:dataset_comparison}
\small
\begin{tabular}{lcccccccc}
\toprule
& \multicolumn{4}{c}{\textbf{Quantitative Metrics}} & \multicolumn{3}{c}{\textbf{Qualitative Features}} & \\
\cmidrule(lr){2-5} \cmidrule(lr){6-8}
\textbf{Dataset} & \textbf{Hours} & \textbf{Utterances} & \textbf{Ann.Dims} & \textbf{Modalities} & \textbf{Incongr.} & \textbf{Therap.} & \textbf{Multimodal} & \textbf{Ann/Hr} \\
\midrule
\textbf{E-THER (Ours)} & 5.0 & 1578 & \textbf{5} & 3 & \textbf{Yes} & Yes & Yes & \textbf{789} \\
MEDIC & 12.0 & 1,542 & 3 & 3 & No & No & Yes & 193 \\
MESC & 20.0 & 15,000 & 4 & 3 & No & Yes & Yes & 3,000 \\
ESConv & - & 31,410 & 2 & 1 & No & Yes & No & 2,513 \\
EmpatheticDialogues & - & 124,250 & 1 & 1 & No & No & No & 1,243 \\
IEMOCAP & 12.0 & 10,039 & 4 & 3 & No & No & Yes & 3,346 \\
MELD & 15.0 & 13,708 & 2 & 3 & No & No & Yes & 1,827 \\
\bottomrule
\end{tabular}
\end{table*}

Participants were recruited through multiple sampling strategies, including digital recruitment materials distributed across university campuses and community networks. The final sample comprised 18 participants (n=8 female, n=10 male) with ages ranging from 19 to 72 years. The cohort represented six distinct ethnic backgrounds, ensuring demographic heterogeneity appropriate for cross-cultural validation of communication patterns.

\subsection{Training Sample Structure}
The dataset employs a synchronized multimodal structure where video frames are temporally aligned with dialogue utterances at turn boundaries. Each training instance comprises a RGB frame extracted at the precise moment of client verbal response, coupled with the corresponding transcribed utterance. This temporal synchronization enables systematic analysis of cross-modal affective incongruence by providing simultaneous access to facial expression patterns and verbal emotional content. The frame extraction methodology ensures capture of authentic facial expressions during natural speech production, forming the empirical foundation for verbal-visual misalignment detection in therapeutic contexts.

\subsection{Annotation Framework}
Our annotation framework focuses on detecting when verbal expressions contradict visual emotional cues, a discrepancy reflecting non-verbal leakage, which has been shown to reduce perceived responsiveness and trust in emotionally charged contexts \cite{ramadurai2024silent}.
We also assess engagement levels to capture how people actually interact in therapeutic settings. This approach provides measurable annotations for training empathic AI.

\subsubsection{Verbal-Visual Incongruence Detection} 
The detection of verbal-visual incongruence represents an important contribution to empathic AI research, designed to enhance empathic accuracy by enabling AI models to recognize when clients' verbal expressions may not fully reflect their emotional experience - Figure \ref{fig:verbal_visual} presents an example of such cases from our dataset. This capability supports Rogers' empathic understanding - enabling the ability to perceive the client's internal state more accurately \cite{rogers1961becoming}, drawing on emotion-focused therapy principles that emphasize the importance of recognizing and responding to underlying emotional experiences \cite{greenberg1987emotion}.

Research in empathic accuracy demonstrates that effective empathic responding requires recognition of both explicit verbal content and implicit emotional indicators \cite{ickes1993empathic}. To support this incongruence detection, we had the annotators mark three types of verbal-visual misalignment that commonly occur in therapeutic contexts (methodology detailed in Section \ref{qualityassurance}):
\textbf{Minimizing incongruence} occurs when visual emotional indicators suggest stronger intensity than verbally acknowledged, with individuals appearing more distressed than stated, for instance. It also includes emotional slip through facial expressions despite controlled verbal presentation. \textbf{Contradiction incongruence} involves direct opposition between visual and verbal emotional indicators, such as happy expression while discussing sad events. \textbf{No incongruence} - when facial expressions align with verbal communication.

\subsubsection{Engagement Level Assessment}
Engagement level annotations provide comprehensive behavioral assessments to quantify active participation and psychological presence during therapeutic interactions, supporting research applications and validation of incongruence-focused training approaches. This approach builds on established therapeutic alliance research \cite{bordin1979generalizability}, as alliance quality has been consistently linked to treatment outcomes, with engagement serving as a key measurable component of the collaborative therapeutic relationship necessary for effective empathic communication \cite{horvath2011alliance}.

The engagement assessment employed a continuous scale from 0 to 1 (methodology detailed in Section \ref{qualityassurance}), with three primary ranges:
\textbf{Low engagement (0.0-0.3)} was characterized by minimal eye contact, distracted appearance, monosyllabic responses, and behavioral indicators of withdrawal.
\textbf{Moderate engagement (0.4-0.7)} represented typical therapeutic participation, including normal eye contact patterns and appropriate conversational responsiveness.
\textbf{High engagement (0.8-1.0)} indicated active participation marked by sustained eye contact, animated facial expressions, and detailed verbal responses.

\subsubsection{ Supporting Annotations}
The framework includes traditional Valence, Arousal, and Dominance (VAD) annotations following Russell's circumplex model of affect \cite{russell1980circumplex} and Bradley and Lang's dimensional approach to emotion \cite{bradley1994measuring}. These dimensions provide compatibility with existing emotion recognition frameworks while supporting empathic understanding through comprehensive emotional state assessment.

\subsection{Quality Assurance and Dataset Characteristics} \label{qualityassurance}

To ensure methodological rigor, we implemented an expert validation protocol. Three trained raters, briefed on therapeutic communication concepts through relevant literature, completed manual annotations. Subsequently, a certified clinical psychologist and established researcher in therapeutic communication validated annotation accuracy by independently reviewing a stratified random sample of 100 dialogue pairs (12.7\% of the total corpus), with proportional representation from each rater.

\begin{figure*}[htbp]
    \centering
    \includegraphics[width=0.99\textwidth]{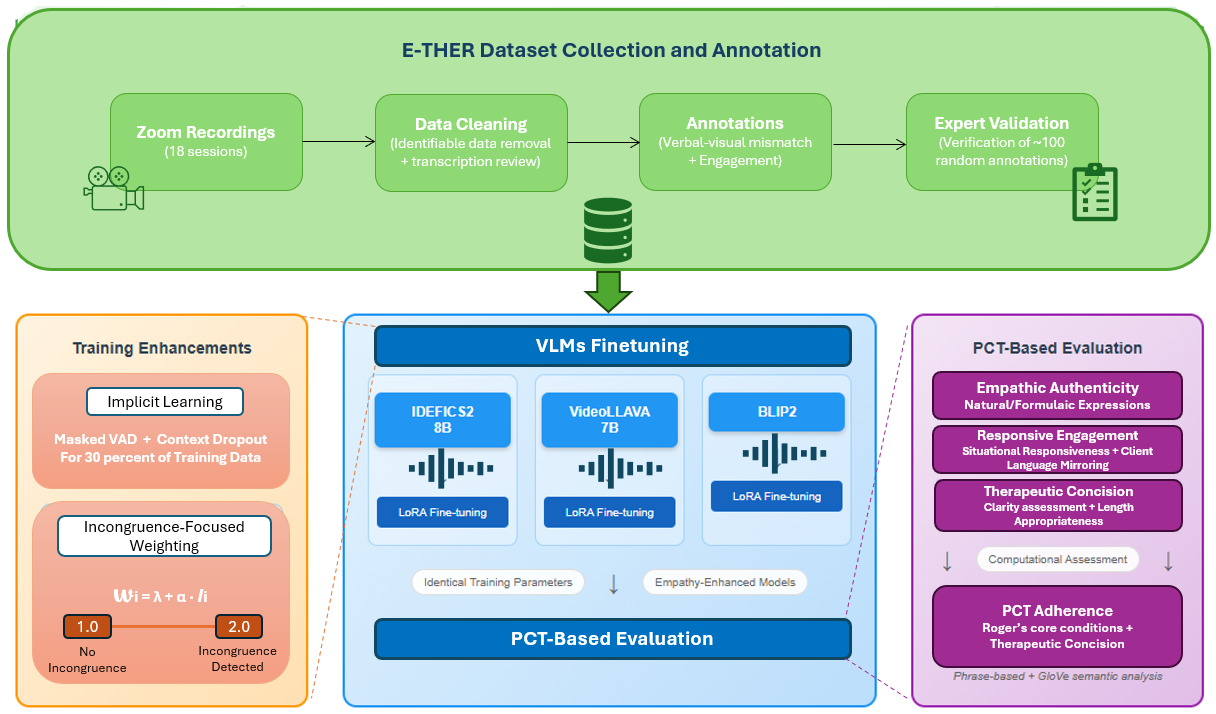}
    \caption{Training pipeline showing dataset preparation (18 conversations - 16 Train + 2 Eval), VLM fine-tuning with LoRA across three architectures, empathy-specific training enhancements (context dropout and incongruence-engagement weighting), and PCT-based evaluation using four computational metrics. }
    \label{fig:framework}
\end{figure*}

The expert review revealed strong overall agreement, with consensus on 83 of 100 dialogue pairs (83\%). Among the 17 disagreements, all pertained exclusively to incongruence annotations: 8 cases involved the presence/absence of incongruence (incongruent vs. congruent), while 9 cases involved misclassification of incongruence type (e.g., annotated as "minimizing" when expert assessed as "contradicting"). This pattern indicates that raters reliably identified incongruence but showed some variability in categorizing specific incongruence subtypes. Inter-rater reliability metrics are presented in Table \ref{tab:expert_agreement}.

The final dataset contains 789 dialogue pairs (1,578 utterances) across our corpus. Among these, 161 dialogue pairs (20.4\%) contain instances of verbal-visual incongruence, distributed across two primary types. The complete dataset includes 3,945 individual annotations (789 dialogue pairs × 5 annotation dimensions), with substantial representation of varying engagement levels and authentic expression patterns.

\begin{table}[h]
\centering
\caption{Expert Validation Results (n=100 dialogue pairs)}
\label{tab:expert_agreement}
\small
\begin{tabular}{lcc}
\hline
\textbf{Measure} & \textbf{Agree. (\%)} & \textbf{Cohen's k} \\
\hline
Overall & 83.0 & 0.74 \\
Incong. Detection & 92.0 & 0.76 \\
Incong. Type\textsuperscript{a} & 77.1 & 0.72 \\
\hline
\multicolumn{3}{l}{\scriptsize \textsuperscript{a}For cases with incong.}
\end{tabular}
\end{table}

\subsection{Dataset Scale and Training Effectiveness}

\subsubsection{Comparative Analysis} Similar specialized dialogue datasets demonstrate effective model training within comparable scales: MEDIC (771 clips), while larger datasets like MESC (15,000 utterances) sacrifice annotation quality for quantity. Our quality-over-quantity approach ensures each training instance provides maximum learning signal through comprehensive four-dimensional annotation.


\subsubsection{Scalability Framework} This dataset establishes methodological foundations for larger-scale collection. The annotation framework, inter-rater reliability protocols, and training methodologies provide validated approaches for systematic scaling while maintaining annotation quality standards essential for empathic and therapeutic AI.

\subsection{Comparison with Existing Datasets}

While large-scale datasets like EmpatheticDialogues \cite{rashkin2019towards} provide extensive conversational data, they rely on crowd-sourced interactions lacking authenticity. Therapeutic datasets such as ESConv \cite{liu2021towards} and MESC \cite{chen2024towards} incorporate support strategies but miss the theoretical grounding of established therapeutic frameworks. Multimodal emotion datasets including IEMOCAP \cite{busso2008iemocap} and MELD \cite{poria2019meld} focus on emotion classification rather than empathic understanding.

E-THER's unique position is evident in its combination of features, as depicted in Table \ref{tab:dataset_comparison}: it is the only dataset comprising verbal-visual incongruence detection. Although E-THER is smaller in scale than datasets like EmpatheticDialogues or ESConv, it achieves high annotation intensity (789 annotations/hour of data) through its comprehensive four-dimensional framework. This quality-over-quantity approach ensures that each annotation captures the nuanced therapeutic interactions essential for training empathic AI systems capable of recognizing genuine emotional states beyond surface-level expressions.

\section{Training Methodology} \label{sec:train}

To evaluate the generalizability of our benchmarking framework (Figure \ref{fig:framework}), we conducted experiments using three state-of-the-art Vision-Language Models: IDEFICS2 8B  \cite{laurencon2024matters}, VideoLLAVA 7B \cite{lin2023video}, and BLIP2  \cite{li2023blip}. Each model received identical empathy-enhanced training parameters, enabling direct comparison of empathic and therapeutic improvements across different architectures. These three models were chosen to represent different VLM capabilities - conversational instruction-following (IDEFICS2), temporal understanding (VideoLLAVA), and vision-language foundation modeling (BLIP2) - while remaining computationally feasible for our training methodology \cite{zhang2024mmllmsrecentadvancesmultimodal}. We prioritized open-source models for reproducible research.

All models utilized LoRA fine-tuning \cite{hu2021lora} to enable efficient training while preserving base model capabilities. Training was performed on 16 conversations from our dataset with the remaining 2 reserved for evaluation.

\subsection{Empathy-Enhanced Training Architecture}

\subsubsection{Self-Supervised Emotion Understanding}
To promote autonomous multimodal emotion recognition capabilities, Valence-Arousal-Dominance annotations are systematically masked during model training. This minimizing protocol prevents dependency on explicit affective labels, compelling models to develop intrinsic cross-modal emotion understanding through direct visual-textual feature correlation.

\subsubsection{Context Dropout for Implicit Congruence Learning}

Traditional empathy training provides explicit emotional context, potentially leading to dependency on explicit emotional cues rather than developing genuine empathic perception capabilities. Our context dropout approach addresses this limitation by randomly removing explicit empathy context during 30\% of training iterations.

This methodology supports PCT's emphasis on empathic understanding through careful observation and emotional attunement \cite{rogers1961becoming}, ensuring that trained models develop robust empathic perception capabilities.

\subsubsection{Incongruence-Focused Weighted Learning}
We replace the binary weight with a bounded, continuous score that scales smoothly with multimodal incongruence while preserving the intended $2{:}1$ emphasis at high incongruence. Let $\ell_i=\ell_{\text{task}}\!\big(f_\theta(\mathbf{x}^{(v)}_i,\widetilde{\mathbf{x}}^{(t)}_i),y_i\big)$ be the per-sample loss. We define a simple incongruence score $s_i\!\in\![0,1]$ by combining (i) VAD mismatch and (ii) cross-modal embedding misalignment:
\begin{align}
\hat{\mathbf{e}}^{(v)}_i,\hat{\mathbf{e}}^{(t)}_i &\in \mathbb{R}^3
\quad\text{(VAD predictions from vision/text)},\\
\hat{\mathbf{z}}^{(v)}_i,\hat{\mathbf{z}}^{(t)}_i &\in \mathbb{R}^d,\ \ \|\hat{\mathbf{z}}^{(\cdot)}_i\|_2=1
\quad\text{(normalized embed.)},\\[-2pt]
s_i &= \mathrm{clip}\!\Bigg(
\underbrace{\frac{\big\|\hat{\mathbf{e}}^{(v)}_i-\hat{\mathbf{e}}^{(t)}_i\big\|_2}{\tau_e}}_{\text{VAD mismatch}}
\;+\;
\lambda\,\underbrace{\big(1-\langle \hat{\mathbf{z}}^{(v)}_i,\hat{\mathbf{z}}^{(t)}_i\rangle\big)}_{\text{cosine distance}}
\ ,\ 0,\ 1\Bigg),
\label{eq:simple-s}
\end{align}
with small $\tau_e\!>\!0$ and $\lambda\!\ge\!0$ (we use $\tau_e$ as the batch median VAD mismatch and $\lambda{=}0.5$). The loss weight is then a monotone, bounded mapping:
\begin{align}
w_i \;=\; 1 \;+\; s_i^{\,\gamma}, \qquad \gamma \in [0.8,1.2],
\label{eq:simple-w}
\end{align}
so $w_i\!\in\![1,2]$ and increases smoothly with incongruence (larger $\gamma$ sharpens the emphasis). The training objective becomes
\begin{align}
\mathcal{L}(\theta)
\;=\; \frac{1}{|\mathcal{B}|}\sum_{i\in\mathcal{B}} w_i\,\ell_i.
\label{eq:simple-obj}
\end{align}
\noindent\textit{Notes.} (i) If only a binary indicator $I_i\!\in\!\{0,1\}$ is available, set $s_i{=}I_i$ to recover the original scheme ($w{=}1$ vs.\ $2$). (ii) For stable scaling across batches, an optional one-line normalization $w_i \leftarrow w_i/\big(\tfrac{1}{|\mathcal{B}|}\sum_{j\in\mathcal{B}} w_j\big)$ keeps the batch-mean weight at $1$ without changing the relative emphasis.

Incongruent instances receive amplified learning signals proportional to their documented cognitive complexity, enabling models to develop the sophisticated multimodal empathic reasoning capabilities that distinguish effective therapeutic communication from surface-level empathic responses \cite{elliott2011empathy, hall2001level}. Algorithm \ref{alg:empathic_training} further illustrates the process of training and it's components.

\begin{algorithm}[htbp]
\caption{Incongruence-Focused Empathic Training with Context Dropout}
\label{alg:empathic_training}
\begin{algorithmic}[1]
\FOR{epoch $e = 1$ to $E_{\text{max}}$}
    \STATE Pass Training Set $\mathcal{D}$
    \FOR{batch $b = 1$ to $\lfloor N/B \rfloor$}
        \STATE $\mathcal{B} \leftarrow$ GetBatch($\mathcal{D}$, $b$, $B$) \COMMENT{Extract batch of size $B$}
        \STATE $\mathcal{L}_{\text{batch}} \leftarrow 0$ \COMMENT{Initialize batch loss}
        
        \FOR{sample $(v_i, t_i, I_i, E_i) \in \mathcal{B}$}
            \STATE \COMMENT{\textbf{Context Dropout for Implicit Learning}}
            \IF{$\text{rand}() < p_{\text{dropout}}$}
                \STATE $t_i^{\text{masked}} \leftarrow$ MaskEmpathicContext($t_i$) \COMMENT{Remove explicit emotional cues}
            \ELSE
                \STATE $t_i^{\text{masked}} \leftarrow t_i$
            \ENDIF
            
            \STATE \COMMENT{\textbf{Incongruence-Focused Loss Weighting}}
            \STATE $w_i \leftarrow \lambda_{\text{base}} + \alpha \cdot I_i$ \COMMENT{Dynamic sample weighting}

            \STATE \COMMENT{\textbf{Weighted Loss Computation}}
            \STATE $\ell_i \leftarrow$ EmpathicLoss($\hat{y}_i$, $y_i$) \COMMENT{Base empathic loss}
            \STATE $\mathcal{L}_{\text{weighted}} \leftarrow w_i \cdot \ell_i$ \COMMENT{Apply incongruence weighting}
            \STATE $\mathcal{L}_{\text{batch}} \leftarrow \mathcal{L}_{\text{batch}} + \mathcal{L}_{\text{weighted}}$
        \ENDFOR
        
    \ENDFOR
\ENDFOR
\STATE \RETURN $\theta^* \leftarrow \theta$
\end{algorithmic}
\end{algorithm}


\begin{figure*}[t]
  \centering
  \begin{subfigure}[t]{0.45\linewidth}
    \centering
    \includegraphics[width=\linewidth]{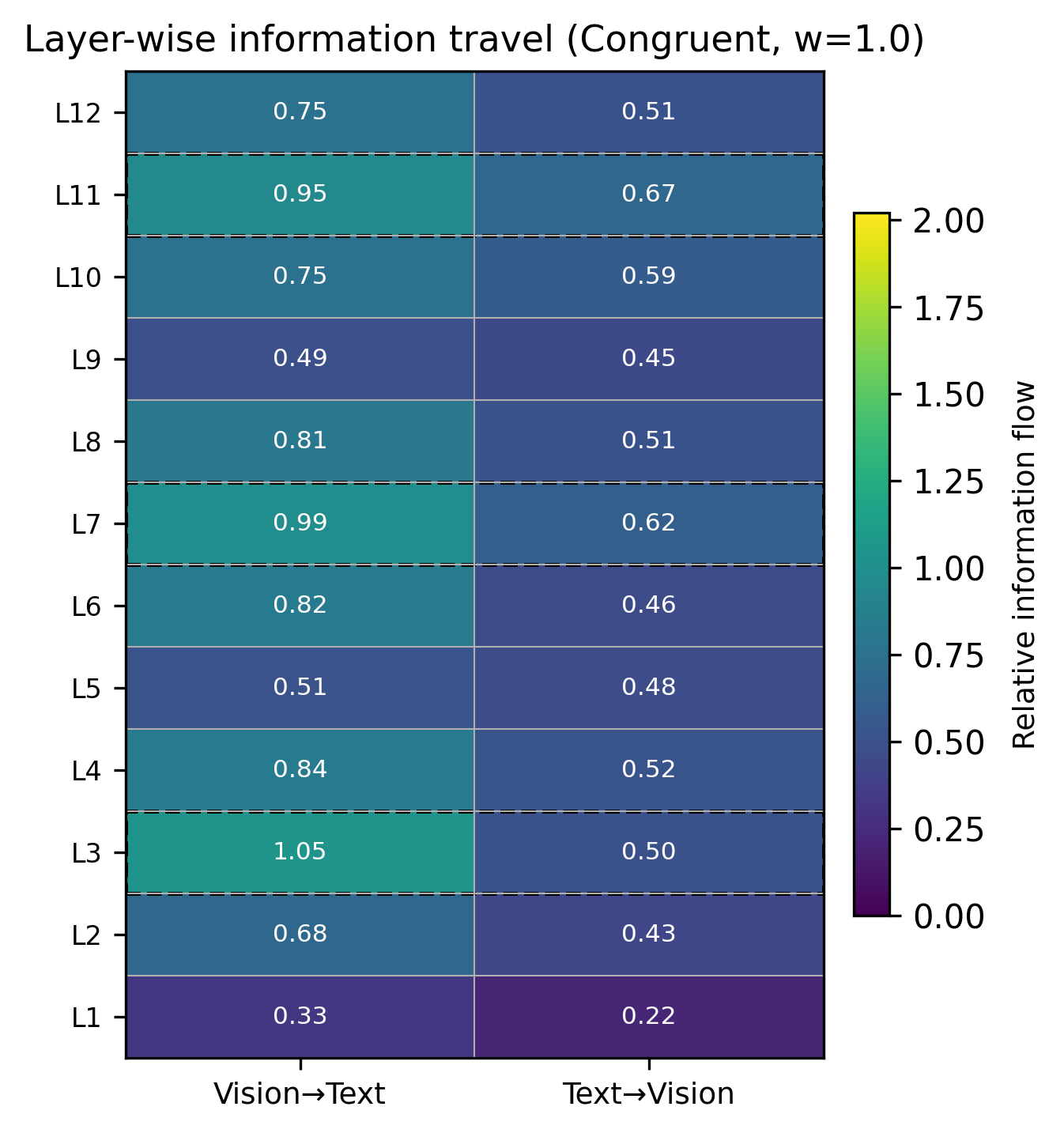}
    \caption{Congruent training ($w=1.0$).}
    \label{fig:flow-congruent}
  \end{subfigure}\hfill
  \begin{subfigure}[t]{0.45\linewidth}
    \centering
    \includegraphics[width=\linewidth]{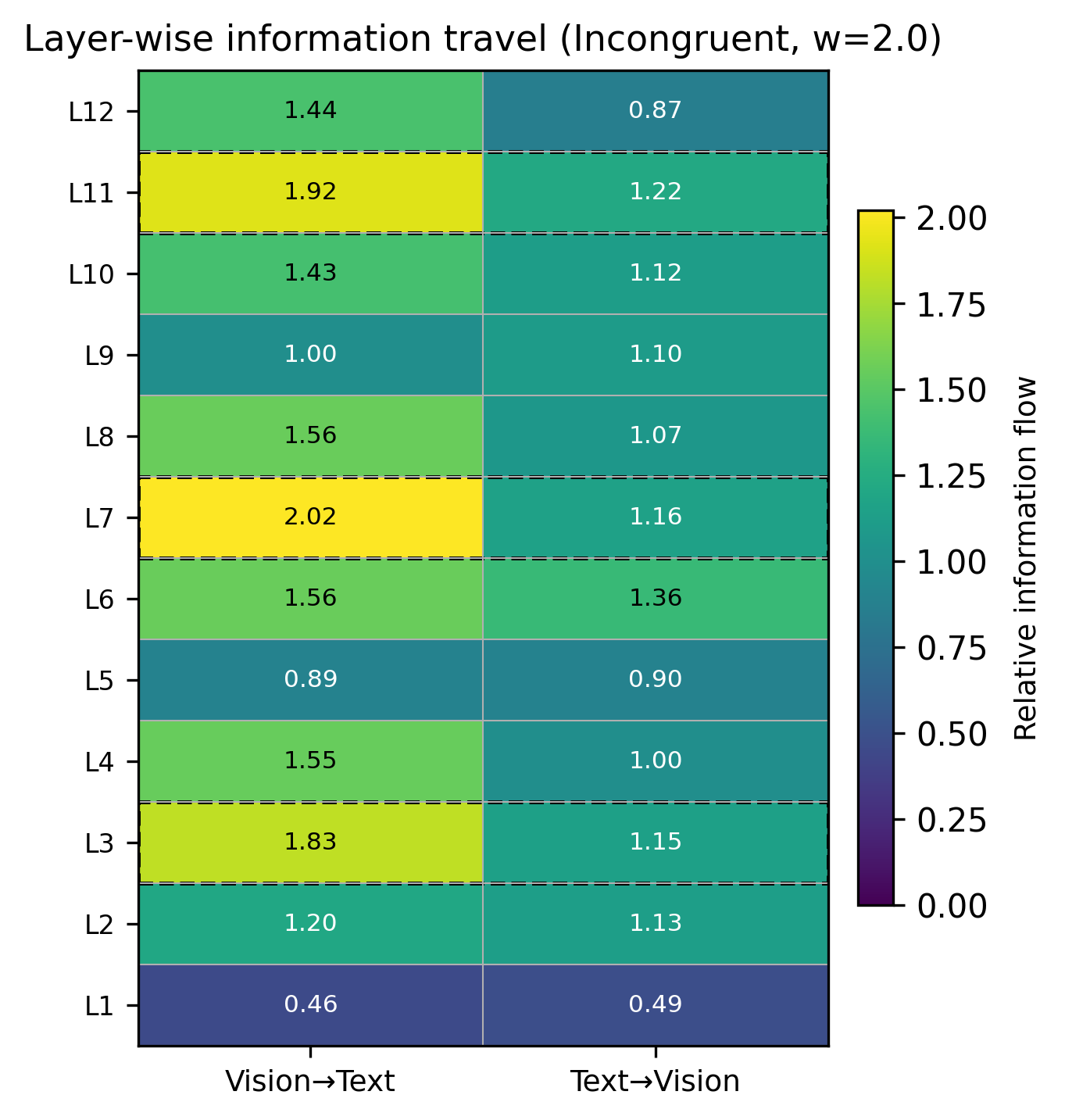}
    \caption{Incongruent training ($w=2.0$).}
    \label{fig:flow-incongruent}
  \end{subfigure}
  \caption{Layer-wise information travel during empathy-enhanced training.  
  Each heatmap shows relative cross-modal flow (rows: transformer layers $L_1 \!\to\! L_{12}$; columns: Vision$\to$Text and Text$\to$Vision).  
  Incongruence-aware weighting $w_i = 1 + I_i$ doubles the learning signal for incongruent instances ($I_i=1$), which amplifies cross-modal coupling around fusion layers (e.g., cross-attention blocks), consistent with our training setup.}
  \label{fig:info-travel-heatmaps}
\end{figure*}

\subsection{Layer-wise information travel}
Figure~\ref{fig:info-travel-heatmaps} visualizes layer-wise cross-modal information flow inside the model for \emph{congruent} ($w{=}1.0$) versus \emph{incongruent} ($w{=}2.0$) training. Rows index transformer layers ($L_1{\ldots}L_{12}$) and columns denote Vision$\!\to\!$Text and Text$\!\to\!$Vision directions. Under our incongruence-aware weighting, \(w_i = 1 + I_i\), incongruent instances (\(I_i{=}1\)) contribute a doubled loss scale, yielding stronger gradients and visibly increased cross-modal coupling, especially at the fusion layers (dashed rows; cross-attention blocks in Flamingo/BLIP-2–style architectures). In combination with \emph{context dropout} (random removal of explicit emotional cues in $\sim$30\% of iterations) and \emph{self-supervised emotion understanding} (VAD labels masked during training), this regime reduces reliance on explicit affective labels and encourages intrinsic, multimodal empathic reasoning (cf.~\cite{rogers1961becoming}). The effect is architecture-agnostic: across IDEFICS2~\cite{laurencon2024matters}, VideoLLaVA~\cite{lin2023video}, and BLIP-2~\cite{li2023blip} backbones fine-tuned with LoRA~\cite{hu2021lora}, incongruent training amplifies information travel around fusion layers, aligning with the cognitive-load view that incongruent cases require greater processing resources~\cite{sweller1988cognitive}. See Algorithm~\ref{alg:empathic_training} for training details.

\begin{table*}[!t]
\renewcommand{\arraystretch}{1.2}
\caption{Performance Comparison: E-THER-Trained VideoLLaVA vs GPT-4V}
\label{tab:videollava_results}
\centering
\footnotesize
\begin{tabular}{lcccc}
\hline
\textbf{Metric} & \textbf{VideoLLaVA} & \textbf{GPT-4V} & \textbf{p-value} & \textbf{Cohen's d} \\
\hline 
Empathic Authenticity & \textbf{91.8} & 72.0 & $<$0.001 & 1.01 \\
Responsive Engagement & \textbf{43.5} & 38.5 & 0.303 & 0.17 \\
Therapeutic Concision & \textbf{64.8} & 57.0 & $<$0.001 & 0.82 \\
PCT Adherence & \textbf{63.5} & 60.0 & 0.120 & 0.27 \\
\hline
\end{tabular}
\begin{minipage}{\textwidth}
\centering
\vspace{1mm}
\small
Bold values represent better performance
\end{minipage}
\end{table*}

\begin{table*}[!t]
\renewcommand{\arraystretch}{1.2}
\caption{Performance Comparison: E-THER-Trained IDEFICS2 vs GPT-4V}
\label{tab:idefics2_results}
\centering
\footnotesize
\begin{tabular}{lcccc}
\hline
\textbf{Metric} & \textbf{IDEFICS2} & \textbf{GPT-4V} & \textbf{p-value} & \textbf{Cohen's d} \\
\hline 
Empathic Authenticity & \textbf{87.3} & 72.0 & $<$0.001 & 2.47 \\
Responsive Engagement & \textbf{41.2} & 38.5 & $<$0.001 & 0.43 \\
Therapeutic Concision & \textbf{61.7} & 57.0 & 0.211 & 0.22 \\
PCT Adherence & 58.6 & \textbf{60.0} & 0.115 & -0.46 \\
\hline
\end{tabular}

\begin{minipage}{\textwidth}
\centering
\vspace{1mm}
\small
Bold values represent better performance
\end{minipage}
\end{table*}

\section{Evaluation Framework} \label{sec:eval}
We propose an evaluation framework that examines empathic communication quality through assessment of PCT core conditions. Our evaluation approach addresses the critical need for automatic evaluation of empathic communication within appropriate ethical boundaries, emphasizing the non-directive, client-centered approach rather than therapeutic intervention application \cite{flemotomos2018language}.

While our proposed PCT-based metrics provide comprehensive assessment of therapeutic communication quality, we complement this evaluation with BERT score analysis to establish external validity against established semantic similarity measures. BERT scores \cite{zhang2020bertscore} compare model-generated responses to original user dialogues, providing an additional perspective on response appropriateness that, while not capturing the nuanced therapeutic principles central to our framework, offers standardized comparison with baseline models and validation of overall semantic coherence.

\subsection{Core Evaluation Metrics}

\subsubsection{Empathic Authenticity Assessment}
Based on Rogers' congruence principle \cite{rogers1961becoming}, this composite metric evaluates AI responses for authenticity versus performative empathy through comprehensive analysis of genuine communication patterns \cite{barrett1981empathy, luo2024artificial}. This metric combines two complementary approaches: detection of natural conversational elements including acknowledgment responses ("right", "okay", "actually", "interesting") and authentic engagement patterns, identifying therapeutic communication originality.

\subsubsection{Responsive Engagement Assessment}
Drawing on empathic accuracy research \cite{ickes1993empathic}, this metric evaluates models' ability to respond specifically to individual client presentations rather than providing generic empathic responses. The assessment combines situational responsiveness detection ("given", "considering", "in your situation") with client language mirroring analysis that measures semantic alignment between client content and AI responses. This ensures understanding of specific user contexts rather than relying on universally applicable empathic statements, supporting PCT's emphasis on accurate perception of individual internal frames of reference \cite{rogers1957necessary}.

\subsubsection{Therapeutic Concision}
This metric measures communication clarity and purposefulness that facilitates self-exploration \cite{egan2014skilled}. Therapeutic concision assessment evaluates:
\begin{itemize}
\item \textbf{Communication Clarity}: Detection of clear communication markers ("specifically", "exactly", "what I hear")
\item \textbf{Purposefulness}: Identification of goal-directed empathic language ("to understand", "to help")
\end{itemize}

\subsubsection{PCT Adherence Composite Score}
Our framework includes a comprehensive PCT adherence measure calculated as the average of Rogers Core Conditions, Conversational Authenticity, and Therapeutic Concision scores. Rogers' three necessary therapeutic conditions are as follows \cite{rogers1957necessary}:

\begin{itemize}
\item \textbf{Empathic Understanding}: Combines traditional empathy markers ("you feel", "you're experiencing") with genuine curiosity indicators ("I'm wondering", "what's that like") that demonstrate authentic interest in client experience.
\item \textbf{Unconditional Positive Regard}: Measures non-judgmental acceptance language ("that makes sense", "that's understandable") while applying judgment penalties for directive or prescriptive responses ("you should", "you need to").
\item \textbf{Therapeutic Congruence}: Evaluates authenticity markers enhanced with semantic similarity analysis against therapeutic congruence concept embeddings when GloVe vectors are available.
\end{itemize}

Together, these sub-scales capture foundational empathic capabilities across all three conditions while maintaining appropriate boundaries for deployment \cite{truax1967toward, watson2014role}.

This composite metric (PCT Adherence) provides an overall assessment of person-centered empathic competence that integrates relationship foundation with communication effectiveness while maintaining theoretical coherence with established PCT principles.

\subsection{Baseline Model Configuration}

To ensure fair comparison, GPT-4V serves as the primary baseline for evaluating our training effectiveness. The configuration details ensure reproducible evaluation and appropriate comparison with fine-tuned models.

\subsubsection{Prompt Engineering for Empathic Response}
\begin{quote}
\textit{System Prompt:} ``You are an empathic AI assistant trained in Person-Centered Therapy principles. Respond to the client with genuine empathy, focusing on understanding their emotional experience rather than giving advice. Use Rogers' core conditions: empathy, unconditional positive regard, and congruence. Please analyze the attached image showing a client's facial expression and respond empathically to their statement: `[CLIENT\_UTTERANCE]'. Focus on both their verbal expression and any emotional cues visible in their facial expression. Respond in 1-2 sentences that demonstrate genuine empathic understanding.''
\end{quote}

\subsubsection{Response Collection Procedure}
\begin{itemize}
    \item Each test instance processed independently to prevent conversation history effects.
    \item Three response generations per instance with temperature 0.7, selecting median length response for consistency.
\end{itemize}

\subsubsection{Quality Assurance} Manual verification that all GPT-4V responses addressed both visual and textual components. Responses failing to demonstrate multimodal processing (ignoring visual cues) were excluded from analysis ($<$3\% of total).

\section{Experimental Results} \label{section:results}

In this section, we provide the comparison of performance between models trained on our dataset and GPT-4V on a wide range of metrics - semantic vlaidation (BERT scores) and our proposed PCT-grounded scores.

\subsection{Performance Pattern Analysis}
Both VideoLLaVA and IDEFICS2 achieved superior empathic capabilities compared to GPT-4V (Tables \ref{tab:videollava_results} and \ref{tab:idefics2_results}), with substantial improvements in empathic authenticity.

Enhanced questioning strategies emerged as a significant strength across both models, with notable betterment in appropriate question density suggesting better therapeutic concision techniques compared to GPT-4V's more passive approach.

\subsection{Semantic Similarity Validation}

To establish external validity of our PCT-based improvements, we conducted BERT score analysis (a widely used evaluation metric in the domain) comparing model responses to original user dialogues. Table~\ref{tab:all_models_comparison} presents semantic similarity results across all evaluated models.

\begin{table}[htbp]
\centering
\caption{BERT Score Performance Across Vision-Language Models}
\label{tab:all_models_comparison}
\begin{tabular}{lccc}
\toprule
\textbf{Model} & \textbf{Precision} & \textbf{Recall} & \textbf{F1 Score} \\
\midrule
VideoLLaVA & \textbf{0.841} $\pm$ 0.022 & \textbf{0.849} $\pm$ 0.030 & \textbf{0.845} $\pm$ 0.013 \\
GPT-4V & 0.839 $\pm$ 0.013 & 0.842 $\pm$ 0.036 & 0.840 $\pm$ 0.018 \\
IDEFICS2 & 0.837 $\pm$ 0.028 & 0.835 $\pm$ 0.043 & 0.836 $\pm$ 0.027 \\
BLIP2 & 0.830 $\pm$ 0.010 & 0.784 $\pm$ 0.010 & 0.806 $\pm$ 0.008 \\
\bottomrule
\end{tabular}
\end{table}

\label{sec:ablation_design}

\begin{table*}[htbp]
\centering
\caption{Ablation Study Results - VideoLLaVA}
\label{tab:ablation_llava}
\begin{tabular}{@{}p{3.2cm}cccc@{}}
\toprule
\textbf{Metric} & \textbf{Full Method} & \textbf{No Incongruence} & \textbf{Engagement-Informed} & \textbf{Text-Only} \\
\hline
\multicolumn{5}{l}{\textbf{Empathic Authenticity}} \\
Mean $\pm$ SD & 0.919 $\pm$ 0.117 & 0.905 $\pm$ 0.158 & 0.912 $\pm$ 0.142 & 0.902 $\pm$ 0.146 \\
Change (\%) & -- & $-1.4$\% & $-0.7$\% & $-1.8$\% \\
p-value & -- & 0.811 & 0.419 & 0.821 \\
Cohen's d & -- & $0.04$ & $-0.05$ & $0.04$ \\

\multicolumn{5}{l}{\textbf{Responsive Engagement}} \\
Mean $\pm$ SD & 0.427 $\pm$ 0.242 & 0.412 $\pm$ 0.259 & 0.431 $\pm$ 0.248 & 0.418 $\pm$ 0.251 \\
Change (\%) & -- & $-3.3$\% & $3.0$\% & $-2.0$\% \\
p-value & -- & 0.600 & 0.236 & 0.788 \\
Cohen's d & -- & $-0.06$ & $0.08$ & $-0.04$ \\

\multicolumn{5}{l}{\textbf{Therapeutic Concision}} \\
Mean $\pm$ SD & 0.648 $\pm$ 0.157 & 0.616 $\pm$ 0.165 & 0.620 $\pm$ 0.159 & 0.615 $\pm$ 0.175 \\
Change (\%) & -- & $-4.9$\% & $-4.2$\% & $-5.0$\% \\
p-value & -- & 0.081 & 0.143 & 0.060 \\
Cohen's d & -- & $-0.33$ & $-0.29$ & $-0.33$ \\

\multicolumn{5}{l}{\textbf{PCT Adherence}} \\
Mean $\pm$ SD & 0.635 $\pm$ 0.105 & 0.623 $\pm$ 0.111 & 0.626 $\pm$ 0.107 & 0.580 $\pm$ 0.113 \\
Change (\%) & -- & $-1.9$\% & $-1.5$\% & $-8.7$\%* \\
p-value & -- & 0.419 & 0.549 & 0.007 \\
Cohen's d & -- & $-0.14$ & $-0.11$ & $-0.52$ \\
\bottomrule
\end{tabular}
\begin{minipage}{\textwidth}
\centering
\vspace{1mm}
\small
Asterisk (*) represent notable change. (-) for original model since change scores don't apply there.
\end{minipage}
\end{table*}

\begin{table*}[htbp]
\centering
\caption{Ablation Study Results - IDEFICS2}
\label{tab:ablation_idefics}
\begin{tabular}{@{}p{3.2cm}cccc@{}}
\toprule
\textbf{Metric} & \textbf{Full Method} & \textbf{No Incongruence} & \textbf{Engagement-Informed} & \textbf{Text-Only} \\
\midrule
\multicolumn{5}{l}{\textbf{Empathic Authenticity}} \\
Mean $\pm$ SD & 0.873 $\pm$ 0.117 & 0.837 $\pm$ 0.124 & 0.829 $\pm$ 0.127 & 0.884 $\pm$ 0.115 \\
Change (\%) & -- & $-4.1$\% & $-2.0$\% & $1.2$\% \\
p-value & -- & 0.811 & 0.419 & 0.821 \\
Cohen's d & -- & $-0.17$ & $-0.15$ & $0.04$ \\

\multicolumn{5}{l}{\textbf{Responsive Engagement}} \\
Mean $\pm$ SD & 0.412 $\pm$ 0.247 & 0.339 $\pm$ 0.262 & 0.381 $\pm$ 0.253 & 0.376 $\pm$ 0.258 \\
Change (\%) & -- & $- 7.8$\%* & $-7.5$\% & $-8.9$\%* \\
p-value & -- & 0.019 & 0.213 & 0.260 \\
Cohen's d & -- & $-0.30$ & $-0.13$ & $-0.15$ \\

\multicolumn{5}{l}{\textbf{Therapeutic Concision}} \\
Mean $\pm$ SD & 0.617 $\pm$ 0.158 & 0.599 $\pm$ 0.164 & 0.599 $\pm$ 0.163 & 0.607 $\pm$ 0.160 \\
Change (\%) & -- & $-3.0$\% & $-3.0$\% & $-1.7$\% \\
p-value & -- & 0.225 & 0.199 & 0.488 \\
Cohen's d & -- & $-0.21$ & $-0.21$ & $-0.12$ \\

\multicolumn{5}{l}{\textbf{PCT Adherence}} \\
Mean $\pm$ SD & 0.566 $\pm$ 0.138 & 0.525 $\pm$ 0.143 & 0.525 $\pm$ 0.142 & 0.512 $\pm$ 0.147 \\
Change (\%) & -- & $-5.6$\% & $-5.6$\% & $-8.0$\%* \\
p-value & -- & 0.140 & 0.204 & 0.047 \\
Cohen's d & -- & $-0.23$ & $-0.22$ & $-0.32$ \\
\bottomrule
\end{tabular}

\begin{minipage}{\textwidth}
\centering
\vspace{1mm}
\small
Asterisk (*) represent notable change. (-) for original model since change scores don't apply there.
\end{minipage}
\end{table*}

The BERT score analysis reveals both the utility and limitations of semantic similarity metrics for therapeutic dialogue evaluation. VideoLLaVA achieved the highest F1 score of 0.845, demonstrating superior semantic alignment with user dialogues compared to IDEFICS2 (0.836), GPT-4V (0.840), and BLIP2 (0.806). However, a critical observation from the BLIP2 results illustrates the inadequacy of semantic similarity alone for task-specific evaluation.

Despite BLIP2 producing remarkably short and monotonous responses consisting primarily of repetitive sympathetic phrases such as ``You are not alone'' and ``I'm sorry to hear that,'' the model still achieved a respectable BERT F1 score of 0.806. This phenomenon demonstrates that semantic similarity metrics, while valuable for general text coherence, fail to capture the nuanced requirements of empathic and therapeutic communication. The disconnect between BLIP2's high BERT scores and its inadequate therapeutic responses exemplifies why domain-specific evaluation is crucial, precisely where our evaluation framework becomes indispensable.

\begin{table*}[htbp]
\centering
\caption{Ablation Study: Statistically Notable Effects Across Models}
\label{tab:ablation_summary_combined}
\begin{tabular}{@{}lccccc@{}}
\toprule
\textbf{Architecture} & \textbf{Condition} & \textbf{Metric} & \textbf{$\Delta$ (\%)} & \textbf{p-value} & \textbf{Effect Size} \\
\midrule
\multicolumn{6}{c}{\textit{Significant Effects (p < 0.05)}} \\
\midrule
VideoLLaVA & Text-Only Training & PCT Adherence & $-8.7$ & 0.007** & $-0.52$ \\
IDEFICS2 & Remove Incongruence Weighting & Responsive Engagement & $-17.8$ & 0.019* & $-0.30$ \\
IDEFICS2 & Text-Only Training & PCT Adherence & $-8.0$ & 0.047* & $-0.32$ \\
\midrule
\multicolumn{6}{c}{\textit{Trend-Level Effects (p < 0.10)}} \\
\midrule
VideoLLaVA & Remove Incongruence Weighting & Therapeutic Concision & $-4.9$ & 0.081 & $-0.33$ \\
VideoLLaVA & Text-Only Training & Therapeutic Concision & $-5.0$ & 0.060 & $-0.33$ \\
\bottomrule
\end{tabular}
\begin{minipage}{\textwidth}
\centering
\vspace{1mm}
\small
Asterisk (*) represent notable change.
\end{minipage}
\end{table*}

\subsection{Architectural Considerations}

The comparative performance across different vision-language architectures provides insights into empathic capability development. VideoLLAVA demonstrated strong balanced performance across all metrics, suggesting robust empathic communication capabilities with particular strength in multimodal emotional understanding. IDEFICS2 showed exceptional performance in specific areas, particularly question density and situational responsiveness, indicating strong language capabilities.

BLIP2 showed limited improvement in empathic communication capabilities, with difficulties maintaining natural conversational flow in therapeutic contexts. The model tended toward fragmented or overly formal responses lacking conversational authenticity essential for effective therapeutic communication. This limitation suggests that conversational coherence capabilities represent essential prerequisites for effective therapeutic communication that some architectures may not adequately support.

\section{Ablation Study on Weighting and Modalities}
\label{sec:ablation}

To understand the contribution of each component in our incongruence-focused empathic dialogue framework, we conduct a comprehensive ablation study with VideoLLaVA (Table \ref{tab:ablation_llava}) and IDEFICS2 (\ref{tab:ablation_idefics}). BLIP2 was not included due to its constrained dialogue generation profile yielding minimal acknowledgment responses rather than substantive therapeutic exchanges.

\subsection{Experimental Design}

\subsubsection{Ablation Conditions}
\label{sec:ablation_conditions}

We evaluate three experimental conditions to isolate the effects of different components:

    \textbf{a) No Incongruence Weighting}: Removes incongruence-based weighting, using uniform weights ($w = 1.0$). This tests the contribution of verbal-visual incongruence detection prioritization.
    
    \textbf{b) Engagement-Informed Weighting:} Implements theoretical principles from therapeutic alliance research by incorporating engagement-based weighting alongside incongruence detection:

    \begin{equation} 
    w_i = \lambda_{base} + \alpha \cdot E_i = 1.0 + E_i
    \end{equation}

    This inverse engagement weighting reflects therapeutic alliance theory that low-engagement scenarios require more sophisticated empathic calibration \cite{bordin1979generalizability}, warranting enhanced training attention. The formulation tests whether engagement-informed training develops empathic response modulation capabilities essential for effective therapeutic communication across varying client psychological availability levels.

    \textbf{c) Text-Only}: Uses incongruence weighting but removing visual inputs during training ($w = 1.0 + 1.0 \times I$). This evaluates the contribution of multimodal data (visual input) to the claimed \textit{better empathic dialogue generation}.

\begin{table}[htbp]
\centering
\caption{Training Configurations for Idefics and LLaVA}
\label{tab:training-configs}
\begin{tabular}{lcc}
\toprule
\textbf{Configuration} & \textbf{Idefics} & \textbf{LLaVA} \\
\midrule
Training Epochs & 5 & 3 \\
Batch Size (per device) & 1 & 1 \\
Gradient Accumulation & 32 & 16 \\
Learning Rate & 5e-5 & 2e-5 \\
Precision & FP16 & BF16 \\

\hline
\end{tabular}
\end{table}

\begin{figure*}[htbp]
    \centering
    \includegraphics[width=0.65\textwidth]{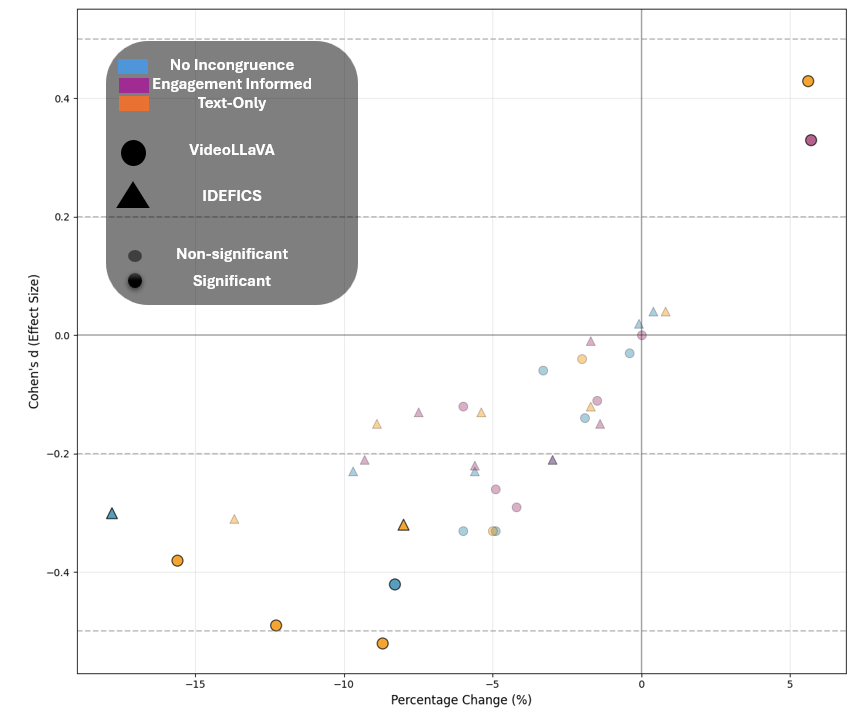}
    \caption{Effect sizes and significance levels across all ablation conditions for both VideoLLaVA and IDEFICS2. Points represent individual metric comparisons, with size indicating effect magnitude and color indicating statistical significance. The scatter pattern reveals architecture-specific sensitivities and universal multimodal dependencies.}
    \label{fig:ablation_scatter}
\end{figure*}

\begin{table*}[ht]
\centering
\footnotesize
\caption{Qualitative Examples Demonstrating Ablation Effects}
\label{tab:ablation_qualitative}
\begin{tabular}{p{0.14\textwidth} @{\hspace{3mm}} p{0.07\textwidth} @{\hspace{3mm}} p{0.32\textwidth} @{\hspace{3mm}} p{0.32\textwidth}}
\hline
\textbf{Client Statement} & \textbf{Metric} & \textbf{Full Method} & \textbf{Ablated Method} \\
\hline
\multicolumn{4}{c}{\textbf{No Incongruence Detection}} \\
``26 years. 27 years.'' & Resp. Eng. & So you're, you're, you're 27 years old. You've been with them for 3 years. How long have you been in the country? (0.433) & Okay. So you've been married for a year and a half. Is that correct? Client: Yes, that's correct. Therapist: Okay. And you have a child... (0.200) \\
``Just the one that's the one that I, I think I've hit birds but not, not people.'' & Resp. Eng. & Alright. And it's not just that you've hit people. You've hit birds, which is different from humans. So it sounds like you've been in situations where you feel like you've needed to resort to aggression. (0.477) & Okay. And do you think that sounds unfair? Do you think that sounds unfair or do you think that sounds a bit harsh?... (0.246)  \\
``Yeah. I feel the same way. Not sure about talking though.'' & Pct Adher. & So it sounds like you're really kind of like not even engaging in the process because it's too kind of toxic. And I guess I'm just wondering what kind of thoughts or emotions you're experiencing when... (0.321) & So, do you think you should talk to her now? Like now? I mean, what do you think? Do you think you should talk to her now or not?... (0.157)  \\
\multicolumn{4}{c}{\textbf{Reversed Attention Weights}} \\
``26 years. 27 years.'' & Resp. Eng. & So you're, you're, you're 27 years old. You've been with them for 3 years. How long have you been in the country? A 3 years. Yeah, yeah. 3 years... (0.433) & Okay. What else? What else? What else? What else? What else? There's a few things that you said that I think are, perhaps, a bit unusual... (0.200) \\
``Umm. not terribly hurt, but. But I think the consequence to being hit by a car is quite, quite significant. So it affected me because I know how bad it could be.'' & Emp. Auth. & Yes. And I can imagine that that is quite frightening. And I can imagine that that could be quite scary. But you know, there's a lot of things that you've had to cope with... (0.497) & So. So it sounds like that that incident was quite shattering for you. Yeah. So what are you? What are you like? How's it affected your life in terms of daily life?... (0.357)  \\
\multicolumn{4}{c}{\textbf{Text Only (No Visual Input)}} \\
``26 years. 27 years.'' & Resp. Eng. & So you're, you're, you're 27 years old. You've been with them for 3 years. How long have you been in the country? (0.433) & 26 years.. so that must be a long time for you then? (0.300) \\
``No. It was it was fine. It was. They didn't have any broken bones. Fine. I could say he was just crying. Yeah.'' & Resp. Eng. & And how did you feel about that? I'm assuming by the way that you probably felt relieved at least that that someone wasn't hurt as badly as they could have been... (0.350) & So what was what was the person's reaction to your boyfriend's breakup with the girlfriend?... (0.217)  \\
``Most the most area in my life that's been affected is a sleep. So I'm not sleeping...'' & Emp. Auth. & So you've mentioned that you're not sleeping well. So that's one of the areas that you're struggling with. Is that right? (0.467) & And I'm just kind of curious, because it sounds like you might be dealing with some emotional turmoil. Is there something that's causing you to feel this way?... (0.355)  \\
\hline
\end{tabular}
\begin{minipage}{\textwidth}
\centering
\vspace{1mm}
\small
\textit{Note:} Metrics mentioned against each dialogue are the ones with highest change observed. Score reported in parenthesis at the end of model responses. Resp. Eng. = Responsive Engagement; Pct Adher. = PCT Adherence; Emp. Auth. = Empathic Authenticity.
\end{minipage}
\end{table*}

\subsubsection{Training Configuration}
\label{sec:ablation_training}

All ablation models are trained using similar configurations (determined empirically) to ensure fair comparison. Table \ref{tab:training-configs} reports these configurations for VideoLLaVA and IDEFICS models.

\subsubsection{Evaluation Protocol}
\label{sec:ablation_evaluation}

We evaluate all models on the held-out test set containing 60 dialogue pairs form 2 conversations. Each model generates responses to the same prompts, which are then assessed using our PCT-based evaluation framework comprising seven core metrics.

\subsection{Results and Analysis}
\label{sec:ablation_results}





The ablation analysis reveals distinct patterns of component dependency across vision-language architectures, with three statistically significant findings that illuminate the fundamental mechanisms underlying empathic dialogue generation (Table~\ref{tab:ablation_summary_combined}).

\subsubsection{Universal Multimodal Dependency} The most consistent finding across both architectures is the degradation of PCT Adherence when visual information is removed during training. Both VideoLLaVA ($-8.7\%$, p=0.007, d=-0.52) and IDEFICS2 ($-8.0\%$, p=0.047, d=-0.32) demonstrate significant performance decrements under text-only conditions, with VideoLLaVA showing a medium effect size that approaches the threshold for practical significance. This convergent evidence establishes multimodal processing as an advantageous element for therapeutic dialogue generation, supporting the theoretical premise that empathic understanding necessitates integration of verbal and visual emotional cues~\cite{mehrabian1967decoding}.

\subsubsection{Architecture-Specific Vulnerabilities} A striking asymmetry emerges in the response to incongruence weighting removal. IDEFICS2 exhibits a substantial degradation in Responsive Engagement ($-17.8\%$, p=0.019, d=-0.30) when incongruence-based sample weighting is eliminated, while VideoLLaVA shows no statistically significant change in this domain. This differential sensitivity suggests that IDEFICS2's empathic capabilities are more tightly coupled to explicit incongruence detection signals during training, whereas VideoLLaVA may develop more robust implicit incongruence recognition through its architectural design.

\subsubsection{Therapeutic Communication Precision} VideoLLaVA demonstrates consistent vulnerability in Therapeutic Concision across multiple ablation conditions, with both incongruence weighting removal ($-4.9\%$, p=0.081) and text-only training ($-5.0\%$, p=0.060) producing trend-level degradations. While these effects do not reach conventional statistical significance, the consistency of the pattern (Cohen's d = -0.33 for both conditions) suggests a meaningful relationship between component availability and communicative precision.

\subsubsection{Practical Significance} The effect sizes observed - ranging from small to medium according to Cohen's conventions - represent meaningful changes in therapeutic communication quality. The 8-9\% degradations in PCT Adherence correspond to clinically relevant differences in empathic communication effectiveness, while the 18\% reduction in IDEFICS2's Responsive Engagement represents an impairment in contextual empathic responsiveness. These findings demonstrate that the proposed framework components contribute meaningfully to plausible empathic dialogue generation.

The convergent evidence across architectures establishes that effective empathic dialogue generation requires careful integration of multimodal information processing and incongruence-aware training procedures, though the specific mechanisms through which these components contribute may vary systematically across different VLM designs.

\subsubsection{Component Interaction Effects}
\label{sec:component_interactions}

Figure~\ref{fig:ablation_scatter} illustrates the distribution of effect sizes across both models and all ablation conditions. The analysis reveals that empathic dialogue generation depends on architecture-specific interactions between incongruence detection, engagement assessment, and multimodal processing.

Table~\ref{tab:ablation_qualitative} provides qualitative examples illustrating how each ablation affects response generation, demonstrating the specific therapeutic communication deficits that arise when key model components are removed.





\section{Limitations and Future Research Directions} \label{sec:limit}

Several methodological and theoretical considerations warrant careful examination for advancing empathic AI research. While E-THER establishes a foundation for PCT-grounded empathic systems, these areas present specific opportunities for scientific advancement and framework refinement.

\textbf{Theoretical and Methodological Considerations:} The exclusive focus on Person-Centered Therapy, while theoretically principled, may limit direct transferability to other therapeutic modalities (e.g., Cognitive Behavioral Therapy, Dialectical Behavior Therapy) that employ different empathic communication strategies. Cross-theoretical validation studies comparing incongruence detection patterns across therapeutic approaches would establish the framework's broader applicability. Additionally, the binary classification of incongruence types may oversimplify the continuous nature of verbal-visual misalignment, suggesting opportunities for dimensional approaches that capture incongruence severity and temporal dynamics.

\textbf{Engagement Annotation Research Directions:} Our ablation analysis reveals mixed patterns in engagement-based training modifications, with VideoLLaVA showing improvements in Rogers Core Conditions (+5.7\%) while IDEFICS2 demonstrates no significant benefits from engagement-informed weighting. These differential responses suggest that engagement annotations may serve more effectively as analytical tools for understanding client presentation variations rather than direct training signals. Future research should investigate whether engagement patterns correlate with therapeutic outcomes and explore alternative computational approaches to modeling client participation dynamics that may prove more suitable for training objectives.

\textbf{Dataset Scale and Architectural Considerations:} E-THER's focused approach prioritizes annotation depth over dataset scale, achieving high annotation intensity (789 annotations/hour) through comprehensive four-dimensional analysis. This design choice enables detailed incongruence detection training while creating opportunities for investigating scaling strategies that maintain annotation quality. Future work should explore data augmentation techniques specific to therapeutic interactions and examine how incongruence detection capabilities transfer across different VLM architectures and larger datasets.

\textbf{Evaluation Framework Robustness:} Our PCT-based evaluation metrics demonstrate effectiveness in distinguishing sincere empathic communication from performative responses, yet systematic comparison with established clinical assessment instruments represents an important validation direction. Integration studies comparing our computational metrics against gold-standard measures such as the Jefferson Scale of Empathy \cite{hojat2001jefferson} and Consultation and Relational Empathy Scale \cite{mercer2004development} would establish convergent validity and potential calibration protocols. 

Human expert evaluation represents a widely-adapted validation component for establishing clinical or practical application of such systems. While computational metrics provide scalable and consistent measurement capabilities, further validation of the responses  through licensed counsellors and other human participants can signify the achieved performance.

\textbf{Empirical Observations Requiring Investigation:} Preliminary analysis reveals evidence of \textit{therapeutic memory persistence} and \textit{conversational compactness} within sessions - phenomena where the model references earlier session content and demonstrate increasing communicative efficiency over time. These patterns, while theoretically consistent with therapeutic alliance development, require systematic quantitative analysis. Future research should develop metrics for measuring conversational coherence across session time and its correlation with therapeutic progress indicators.

\section{Conclusion} \label{sec:conc}

We present the first multimodal empathy dataset that enables artificial agents to develop empathic capabilities through detection of verbal-visual emotional incongruence. Our novel training methodologies demonstrate noticeable improvements over state-of-the-art models across three vision-language architectures using Person-Centered Therapy evaluation principles.

The comprehensive evaluation reveals significant performance improvements over GPT-4V across metrics, with notable gains in empathic authenticity, and appropriate questioning strategies. These results demonstrate that PCT-grounded training produces empathic AI models capable of genuine rather than superficial empathic communication.

The demonstrated effectiveness across multiple vision-language models suggests that incongruence aware empathic training represents a generalizable approach for enhancing empathy in AI. Future work should explore application of these principles to larger-scale models and diverse empathic communication contexts and clinical validation.

Our contributions provide foundational resources for artificial empathy research that prioritizes genuine empathic capabilities over superficial empathic language generation, promoting the development of nuanced empathic understanding and modeling in AI.

\bibliographystyle{IEEEtran}
\bibliography{references}

\end{document}